\documentclass[12pt]{article}
\usepackage{epsfig}
\usepackage{axodraw}
\usepackage{epsfig}                            
\usepackage{graphicx}
\usepackage{rotate}
\usepackage{latexsym}
\newcommand\pubnumber{}
\newcommand\pubdate{\today}
\newcommand\hepnumber{hep-ph/0101139}

\def\csumb{Dipartimento di Fisica Teorica, Universit\`a di Torino, Italy\\
INFN, Sezione di Torino, Italy}
\def\support{\footnote{Work supported by the
European Union under contract HPRN-CT-2000-00149.}} 

\def\Title#1{\begin{center} {\Large\bf #1 } \end{center}}
\def\Author#1{\begin{center}{ \sc #1} \end{center}}
\def\Address#1{\begin{center}{ \it #1} \end{center}}

\newcommand\pubblock{\rightline{\begin{tabular}{l} \pubnumber\\
         \pubdate\\ \hepnumber \end{tabular}}}
\newenvironment{Abstract}{\begin{quotation}  }{\end{quotation}}
\newenvironment{Presented}{\begin{quotation} \begin{center} 
             Presented at the\end{center}
      \begin{center}\begin{large}}{\end{large}\end{center} \end{quotation}}
\def\Acknowledgments{\bigskip  \bigskip \begin{center}
          \large\bf Acknowledgments\end{center}}

\makeatletter
\def\section{\@startsection{section}{0}{\z@}{5.5ex plus .5ex minus
 1.5ex}{2.3ex plus .2ex}{\large\bf}}
\def\subsection{\@startsection{subsection}{1}{\z@}{3.5ex plus .5ex minus
 1.5ex}{1.3ex plus .2ex}{\normalsize\bf}}
\def\subsubsection{\@startsection{subsubsection}{2}{\z@}{-3.5ex plus
-1ex minus  -.2ex}{2.3ex plus .2ex}{\normalsize\sl}}

\renewcommand{\@makecaption}[2]{%
   \vskip 10pt
   \setbox\@tempboxa\hbox{\small #1: #2}
   \ifdim \wd\@tempboxa >\hsize     
       \small #1: #2\par          
     \else                        
       \hbox to\hsize{\hfil\box\@tempboxa\hfil}
   \fi}

 \def\citenum#1{{\def\@cite##1##2{##1}\cite{#1}}}
\def\citea#1{\@cite{#1}{}}
 
\newcount\@tempcntc
\def\@citex[#1]#2{\if@filesw\immediate\write\@auxout{\string\citation{#2}}\fi
  \@tempcnta\z@\@tempcntb\m@ne\def\@citea{}\@cite{\@for\@citeb:=#2\do
    {\@ifundefined
       {b@\@citeb}{\@citeo\@tempcntb\m@ne\@citea\def\@citea{,}{\bf ?}\@warning
       {Citation `\@citeb' on page \thepage \space undefined}}%
    {\setbox\z@\hbox{\global\@tempcntc0\csname b@\@citeb\endcsname\relax}%
     \ifnum\@tempcntc=\z@ \@citeo\@tempcntb\m@ne
       \@citea\def\@citea{,}\hbox{\csname b@\@citeb\endcsname}%
     \else
      \advance\@tempcntb\@ne
      \ifnum\@tempcntb=\@tempcntc
      \else\advance\@tempcntb\m@ne\@citeo
      \@tempcnta\@tempcntc\@tempcntb\@tempcntc\fi\fi}}\@citeo}{#1}}
\def\@citeo{\ifnum\@tempcnta>\@tempcntb\else\@citea\def\@citea{,}%
  \ifnum\@tempcnta=\@tempcntb\the\@tempcnta\else
  {\advance\@tempcnta\@ne\ifnum\@tempcnta=\@tempcntb \else\def\@citea{--}\fi
    \advance\@tempcnta\m@ne\the\@tempcnta\@citea\the\@tempcntb}\fi\fi}
\makeatother

\newcommand{\spro}[2]{{#1}\cdot{#2}}
\newcommand{\ds}{\displaystyle}
\newcommand{\Ddrh}{{\ds\frac{1}{\hat{\varepsilon}}}}

\newcommand{\DiagramFermionToBosonPropagator}[4][85]{
  \vcenter{\hbox{
  \SetScale{0.8}
  \begin{picture}(#1,50)(15,15)
    \ArrowLine(25,25)(50,50)
    \ArrowLine(50,50)(25,75)
    \Photon(50,50)(105,50){2}{8}   \Text(90,40)[lc]{#2}
    \Vertex(50,50){0.5}         \Text(80,48)[cb]{#3}
    \GCirc(82,50){8}{1}            \Text(55,48)[cb]{#4}
    \Vertex(105,50){2}
  \end{picture}}}
  }

\newcommand{\DiagramFermionToBosonFull}[3][70]{
  \vcenter{\hbox{
  \SetScale{0.8}
  \begin{picture}(#1,50)(15,15)
    \ArrowLine(25,25)(50,50)
    \ArrowLine(50,50)(25,75)
    \Photon(50,50)(90,50){2}{8}   \Text(80,40)[lc]{#2}
    \Vertex(50,50){2.5}          \Text(60,48)[cb]{#3}
    \Vertex(90,50){2}
  \end{picture}}}
  }

\newcommand{\bec}{\begin{center}}
\newcommand{\eec}{\end{center}}
\newcommand{\vj}[4]{{\sl #1~}{\bf #2 }\ifnum#3<100 (19#3) \else (#3) \fi #4}
\newcommand{\ej}[3]{{\bf #1~}\ifnum#2<100 (19#2) \else (#2) \fi #3}

\newcommand{\ben}{\begin{enumerate}}
\newcommand{\een}{\end{enumerate}}
\newcommand{\asums}[1]{\sum_{#1}}
\newcommand{\bei}{\begin{itemize}}
\newcommand{\eei}{\end{itemize}}

\newcommand{\bqas}{\begin{eqnarray*}}
\newcommand{\eqas}{\end{eqnarray*}}

\newcommand{\sW}{p_{_W}}
\newcommand{\sZ}{p_{_Z}}

\newcommand{\stws}{s_{\theta}^2}

\newcommand{\drii}[2]{\delta_{#1#2}}                    

\newcommand{\me}{m_e}

\newcommand{\fig}[1]{Fig.~\ref{#1}}

\newcommand{\bq}{\begin{equation}}                   
\newcommand{\eq}{\end{equation}}
\newcommand{\bqa}{\begin{eqnarray}}
\newcommand{\eqa}{\end{eqnarray}}
\newcommand{\ba}[1]{\begin{array}{#1}}
\newcommand{\ea}{\end{array}}
\newcommand{\lpar}{\left(}                            
\newcommand{\rpar}{\right)}

\newcommand{\lcbr}{\left\{}
\newcommand{\rcbr}{\right\}} 
\newcommand{\ph}{\gamma}
\newcommand{\zb}{Z}
\newcommand{\wb}{W}

\newcommand{\barf}{\overline f}

\newcommand{\barnu}{\overline{\nu}}

\newcommand{\mws}{M^2_{_W}}

\newcommand{\mz}{M_{_Z}}
\newcommand{\mzs}{M^2_{_Z}}

\newcommand{\bzms}{M^2_{_0}}

\newcommand{\mes}{m^2_e}

\newcommand{\gf}{G_{\ssF}}
\newcommand{\ssZ}{{\scriptscriptstyle{\zb}}}

\newcommand{\ssF}{{\scriptscriptstyle{F}}}

\newcommand{\ord}[1]{{\cal O}\lpar#1\rpar}

\def\beq{\begin{equation}}
\def\eeq{\end{equation}}
\def\beqar{\begin{eqnarray}}
\def\eeqar{\end{eqnarray}}
\def\barr#1{\begin{array}{#1}}
\def\earr{\end{array}}
\def\bfi{\begin{figure}}
\def\efi{\end{figure}}
\def\btab{\begin{table}}
\def\etab{\end{table}}
\def\bce{\begin{center}}
\def\ece{\end{center}}

\def\nl{\nonumber\\}

\def\mathswitchr#1{\relax\ifmmode{\mathrm{#1}}\else$\mathrm{#1}$\fi}

\def\mathswitch#1{\relax\ifmmode#1\else$#1$\fi}

\newcommand{\bfig}{\begin{center}\begin{picture}}
\newcommand{\efig}[1]{\end{picture}\\{\small #1}\end{center}}

\newcommand{\bmip}[2]{\begin{minipage}[t]{#1pt}\bfig(#1,#2)}
\newcommand{\emip}[1]{\efig{#1}\end{minipage}}

\newcommand{\beanon}{\begin{eqnarray*}}
\newcommand{\eeanon}{\end{eqnarray*}}

\renewcommand{\to}{\rightarrow}

\def\alf1{ {\alpha\over\pi} }

\begin{document}
\begin{titlepage}
\pubblock

\vfill
\def\thefootnote{\fnsymbol{footnote}}
\Title{Non-Annihilation Processes, Fermion-Loop \\[5pt] and QED Radiation}
\vfill
\Author{Giampiero Passarino\support}
\Address{\csumb}
\vfill
\begin{Abstract}
The bulk of large radiative corrections to any process can be obtained
by promoting coupling constants to be running ones and by including QED radiation
at the leading logarithmic level via structure functions evoluted at some scale.
The problem of {\em fixing} the proper scale in running coupling constants and in
structure functions for non-annihilation processes is briefly addressed and
the general solution is analyzed.
\end{Abstract}
\vfill
\begin{Presented}
5th International Symposium on Radiative Corrections \\ 
(RADCOR--2000) \\[4pt]
Carmel CA, USA, 11--15 September, 2000
\end{Presented}
\vfill
\end{titlepage}
\def\thefootnote{\arabic{footnote}}
\setcounter{footnote}{0}

\section{Introduction}

At the eve of LEP shutdown it is of some importance to summarize the present 
status of high precision physics~\cite{Bardin:1999gt}. 
For $e^+e^- \to \barf f$ all one-loop terms are known, including re-summation
of leading terms. At the two-loop level leading and next-to-leading terms
have been computed and included in codes like 
TOPAZ0~\citea{\citenum{Montagna:1993ai},\citenum{Montagna:1993py}}
and ZFITTER~\cite{Bardin:1999yd}.
For realistic observables initial state QED radiation is included via the 
structure function method, or equivalent ones. Final state QED is also 
available as well as the interference between initial and final 
states~\cite{Jadach:2000vf}.
Fine points in QED for $2\to2$ are as follows.
For $s$-channel all the $\ord{\alpha^2L^n},\, n=0,1,2\,\,$ terms are known
from explicit calculations, the  leading $\ord{\alpha^3L^3}$ is also available 
and they are important for the studies of the $\zb$ lineshape.

Differences and uncertainties amount to at most $\pm 0.1\,$MeV on
$\mz$ and $\Gamma_{\ssZ}$ and $\pm 0.01\%$ on 
$\sigma^0_{\rm h}$ (MIZA, TOPAZ0 and ZFITTER)~\cite{ewwg}
For non-annihilation processes (Bhabha) both structure-function
and parton-shower methods have been analyzed and the 
uncertainty is estimated to be $0.061\%$ from BHLUMI~\cite{Jadach:1997is}.
Certainly, full two-loop electroweak corrections are needed for GigaZ 
($10^9 \zb$ events) with a quest for a fast numerical evaluation of the 
relevant diagrams.

For $e^+e^- \to 4\,$fermions all tree-level processes are available and
$\ord{\alpha}$ electroweak corrections are known only for the $\wb\wb$-signal
and in double-pole approximation (DPA)~\cite{Denner:2000bj} 
and~\cite{Jadach:2000kw}.
$e^+e^- \to 4{\rm f} + \ph$ in Born approximation is also available for all 
processes~\cite{Grunewald:2000ju}. 

Fine points in QED for $2\to4$ are as follows
For $e^+e^- \to \wb\wb \to 4\,{\rm f}$ DPA gives the answer but,
for a generic process $e^+e^- \to 4\,{\rm f}$ QED radiation is included
by using $s$-channel structure-functions, \ i.e.
in leading-log approximation.
The latter are strictly applicable only if ISR can be separated unambiguously.
Otherwise their implementation may lead to an excess of radiation.
Preliminar investigations towards non--$s$ SF by GRACE and by 
SWAP~\cite{Grunewald:2000ju} gives an indication on how to implement
the bulk of the non-annihilation effect but still represent {\em ad hoc}
solutions. These methods, which are essentially based on a matching with the
soft photon emission, still contain an ambiguity on the energy scale 
selection with consequences on the predicted observables.

\section{Non-Annihilation processes}

There are several processes, namely those with $t$-channel photons
that are not dominated by annihilation. Typical examples are
single-$\wb$ production and two-photon processes.
The main question can be summarized as follows:
how to include the bulk of radiative corrections?

At the Born level we still require the notion of input parameter set (IPS,
i.e.\  the choice of some set of input parameters (improperly called 
renormalization scheme (RS) in the literature) and of certain relations among 
them, e.g.\
\bqa
\stws &=& 1 - \frac{\mws}{\mzs}, \qquad
\alpha \equiv \alpha_{\gf}= 4 \sqrt{2}\,\frac{\gf\mws\stws}{4\,\pi},
\eqa
Roughly speaking the theoretical uncertainty associated with the choice of the 
RS is most severe whenever low-$q^2$ photons dominate.

The first step in getting the right scales is represented by the Complex-Mass 
Renormalization in the Fermion-Loop approximation which 
gives~\cite{Beenakker:1997kn}
\bqas
\mbox{Couplings}  \quad&\Longrightarrow&\quad  \mbox{Running\,Couplings}  \nl
{} &{}& {}  \nl
\mbox{Transitions} \quad&\Longrightarrow&\quad 
\mbox{Diagonal \, Propagator-Functions} \nl
\eqas
showing a pole in the 2nd sheet, and
\bqas
\mbox{Born \, Vertices} \quad&\Longrightarrow&\quad
\mbox{one \, fermion-loop \, corrected \, Vertices}
\eqas
A typical example is shown by the following identities among diagrams:
\bqas
\DiagramFermionToBosonFull{}{$\ph$} & = &
             \DiagramFermionToBosonPropagator{}{$\ph$}{$\ph$}
           + \DiagramFermionToBosonPropagator[75]{}{$\ph$}{$\zb$},
\nl
\quad
\DiagramFermionToBosonFull{}{$\zb$} & = &
             \DiagramFermionToBosonPropagator{}{$\zb$}{$\ph$}
           + \DiagramFermionToBosonPropagator[75]{}{$\zb$}{$\zb$}.
\label{folldia}
\eqas
Here open circles denote re-summed propagators and the dot a vertex. 
\begin{figure}[t]
\begin{minipage}[t]{14cm}
{\begin{center}
\vspace*{-1.5cm}
\hspace*{-1.0cm}
\mbox{\epsfysize=19cm\epsfxsize=17cm\epsffile{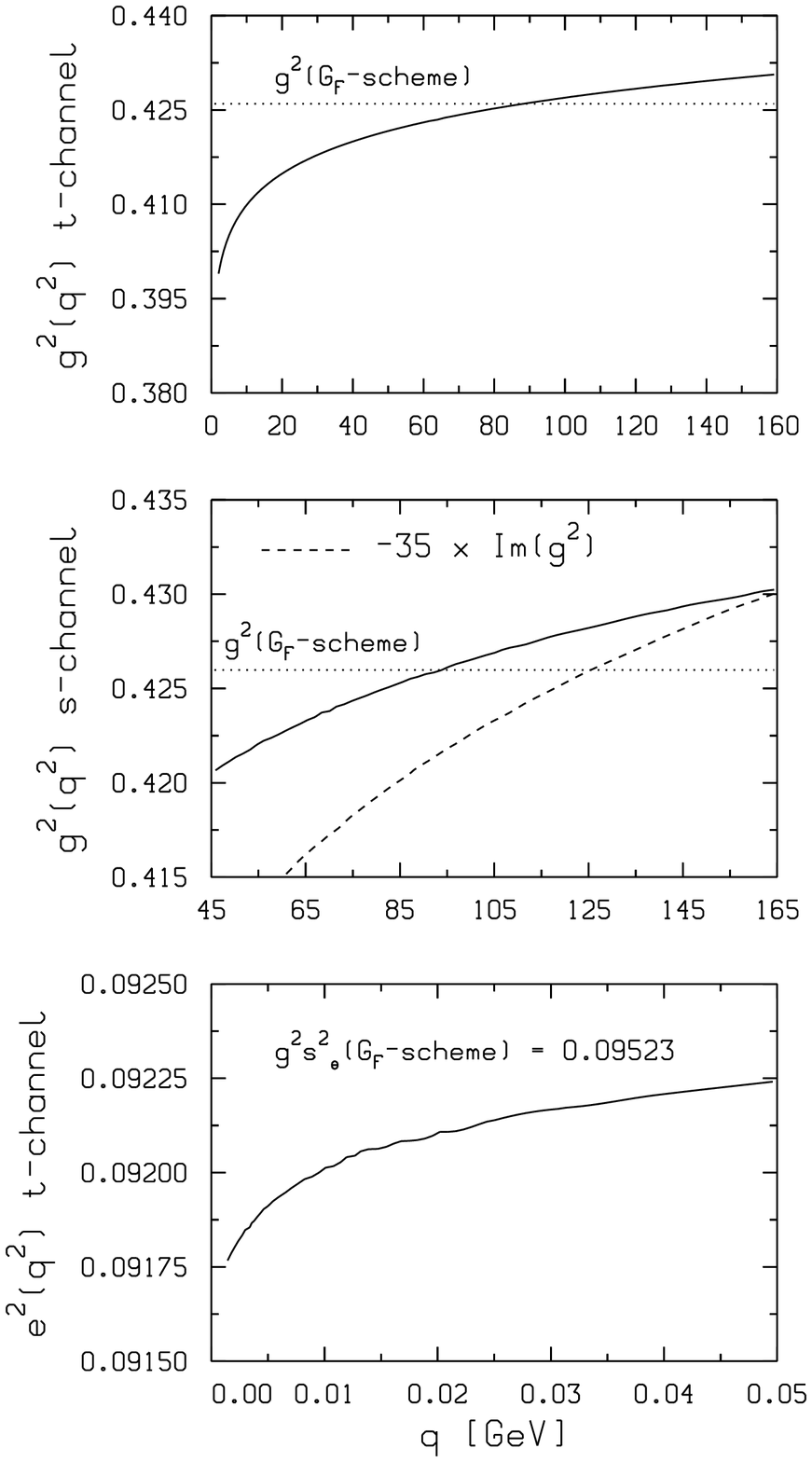}}
\vspace*{-2.7cm}
\end{center}}
\end{minipage}
\label{fig:fig1}
\end{figure}

Running of coupling constants is shown in \fig{fig:fig1}.
In \fig{fig:fig1} the running of $e^2(q^2)$ is
shown for $q^2 \to  0_+$, compared with the fixed value in the $\gf$-scheme.
Furthermore, the evolution of $g^2(q^2)$ is shown for $q^2$ time-like or 
space-like.

The sizeable difference that one gets between $e^2$ running in $t$channel and
$e^2$ fixed in the $\gf$-scheme is one of the major improvements
induced by the FL-scheme in non-annihilation, Born processes.

However, the original formulation of the FL-scheme works only for conserved
external currents.
The extension to external massive fermions exists~\cite{Passarino:2000zh}
and requires one additional replacement: one perform the calculation in the 
$\xi = 1$ gauge, neglects contributions from unphysical scalars and uses
\bqa  
\drii{\mu}{\nu} \, \mbox{(in propagators)} 
\quad&\Longrightarrow&\quad
\drii{\mu}{\nu} + {{p_{\mu}p_{\nu}}\over {M^2(p^2)}},
\eqa
where $M(p^2)$ is the (complex) running mass.
The connection with complex-poles, $\sW,\sZ$ (here only for a massless internal 
world) is simple
\bqa
\wb \quad&\Longrightarrow&\quad
M^2(p^2) = \frac{g^2(p^2)}{g^2(\sW)}\,\sW,  \nl
\zb \quad&\Longrightarrow&\quad
{1\over {\bzms(p^2)}} = \frac{g^2(\sZ)}{g^2(p^2)}\,
\frac{c^2(p^2)}{c^2(\sZ)}\,\frac{1}{\sZ}  \nl
{} \quad&\Longrightarrow&\quad 
{1\over {\bzms(p^2)}} = {{c^2(p^2)}\over {M^2(p^2)}}
\eqa
and gives the M(assive)FL-scheme, where gauge invariance is respected
and collinear regions, e.g.\ outgoing electrons at zero scattering angle,
are accessible for safe theoretical predictions.

\section{Applications to single-$\wb$}
The single-$\wb$ production mechanism is represented in the following figure.
\vspace{0.5cm}
\bqas
\ba{ccc}
\vcenter{\hbox{
  \begin{picture}(110,100)(0,0)
  \ArrowLine(50,50)(0,100)
  \ArrowLine(100,100)(50,50)
  \ArrowLine(0,0)(50,50)
  \ArrowLine(50,50)(100,0)
  \ArrowLine(100,70)(50,50)
  \ArrowLine(50,50)(100,30)
  \GCirc(50,50){15}{0.5}
  \Text(10,100)[lc]{$e^+$}
  \Text(10,0)[lc]{$e^-$}
  \Text(110,100)[lc]{$\barnu_e$}
  \Text(110,70)[lc]{$\barf_2$}
  \Text(110,30)[lc]{$f_1$}
  \Text(110,0)[lc]{$e^-$}
  \end{picture}}}
&\quad+&
\vcenter{\hbox{
  \begin{picture}(110,100)(0,0)
  \ArrowLine(50,50)(0,100)
  \ArrowLine(100,100)(50,50)
  \ArrowLine(100,70)(50,50)
  \ArrowLine(50,50)(100,30)
  \Photon(50,50)(50,10){2}{7}
  \ArrowLine(0,0)(50,10)
  \ArrowLine(50,10)(100,0)
  \GCirc(50,50){15}{0.5}
  \Text(10,100)[lc]{$e^+$}
  \Text(10,0)[lc]{$e^-$}
  \Text(110,100)[lc]{$\barnu_e$}
  \Text(110,70)[lc]{$\barf_2$}
  \Text(110,30)[lc]{$f_1$}
  \Text(110,0)[lc]{$e^-$}
  \Text(60,20)[lc]{$Q^2$}
  \end{picture}}}
\ea
\eqas
\vspace{0.4cm}
\begin{center}
The CC20 family of diagrams with the explicit component\\ 
containing a $t$-channel photon.
\end{center}
\vspace{0.3cm}
The main consequences of applying the MFL-scheme are as follows:

\bei

\item there is a maximal decrease of about $7\%$ in the result if we compare 
with the $\gf$-scheme predictions but, 

\item the effect is rather sensitive to the relative weight of 
multi-peripheral contributions and is process and cut 
dependent~\cite{Passarino:2000mt}. 

\eei

\section{QED radiation for arbitrary processes}

Here the relevant question can be formulated as follows: is 
multi-photon radiation a one-scale or a multi-scale convolution phenomenon?
\bqa
\sigma\lpar p_+p_- \to q_1\dots q_n + \mbox{QED}\rpar &\stackrel{?}{=}&
\int\,dx_+\,dx_-\,D(x_+,?)D(x_-,?) \nl
{} &\times& \sigma\lpar x_+p_+ x_-p_- \to q_1\dots q_n\rpar
\eqa
In the above equation the question mark means that the corresponding scale
has to be guessed. We need to understand how the standard SF-method is 
related to the exact YFS exponentiation.
In the standard YFS treatment of multiple photon emission we have
\bqa
\sigma\lpar p_++p_- \to \asums{i=1,2l}q_i + \asums{j=1,n}k_j\rpar &\sim& 
\int\,dPS_q \mid M_0\mid^2\,
E\lpar p_++p_--\asums{i}q_i\rpar,
\eqa
where $E$ is the spectral function defined by
\bqa
E(K) &=& \frac{1}{(2\,\pi)^4}\,\int\,d^4x\,\exp(i\spro{K}{x})\,E(x),  \nl
E(x) &=& \exp\lcbr \frac{\alpha}{2\,\pi^2}\,\int d^4k e^{i\spro{k}{x}}
\delta^+(k^2)\,\mid j^{\mu}(k)\mid^2\rcbr
\eqa
At this point we choose an alternative procedure were we do not separate the 
soft component from the hard one and compute some exact result valid for an 
arbitrary number of dimensions $n$ and for on-shell photons, i.e. $k^2=0$,
\bqa
I &=& \int d^nk\,e^{i\spro{k}{x}}\,{{\delta^+(k^2)}\over 
{\spro{p_i}{k}\,\spro{p_j}{k}}}
\eqa
In dimensional-regularization one has the following result, valid 
$\forall x^2$:
\bqa
I(x) &=& -\,\pi\,\rho\,\int_{_0}^{^1} \frac{du}{P^2}
\lpar \Ddrh + 2\,\ln 2 - \ln x^2 - \xi\,\ln\frac{\xi+1}{\xi-1}\rpar,  
\eqa
where we have defined a variable $\xi$ as the ratio
\bqa
\xi &=& {{|x_0|}\over r}, 
\eqa
with an infinitesimal imaginary part attributed to $x_0$,
\bqa
x_0 \to x_0 + i\delta. \qquad \delta \to 0_+.
\eqa
Furthermore, $P$ is the linear combination
\bqa
P &=& p_j + \lpar \rho p_i - p_j\rpar\,u,  \nl
\eqa
where we have defined $\rho$ to satisfy
\bqa
\lpar \rho p_i - p_j\rpar^2 &=& 0, 
\eqa
and $x_0, r$ are rewritten in covariant form as follows:
\bqa
x_0 = - {{\spro{P}{x}}\over {\sqrt{-P^2}}},  \quad
r^2 = x^2_0 + x^2.
\eqa
The last integral shows the infrared pole $\Ddrh$ and a collection of 
${\rm Li}_2$-functions. Therefore, $E(K)$ is not available in close form. 
The scheme that we want to propose defines a coplanar 
approximation~\cite{Chahine} to the exact spectral function,
\bqa
I^c_{ij} &\stackrel{\rm def}{=}& 
- \frac{2}{3}\,\pi \rho_{ij}\,{\cal F}_{\rm cp}\,
{1\over {p^2_j-\rho^2_{ij}\,p^2_i}}\,\ln\frac{\rho^2_{ij}p^2_i}{p^2_j}, \nl
I^c_{ii} &\stackrel{\rm def}{=}& 
- \frac{2}{3}\,\pi \rho_{ij}\,{\cal F}_{\rm cp}\,\frac{1}{m^2_i},
\nl
{\cal F}_{\rm cp} &=& \ln\lcbr e^{-\Delta_{\rm IR}} \,\, {{\spro{p_i}{x}\,
\spro{p_j}{x}}\over {m_im_j}}\rcbr,  \nl
\Delta_{\rm IR} &=& \Ddrh + {\rm constants}.
\eqa
Within the coplanar approximation we have
\bqa
E^{{\rm pair}\,<ij>}(K) &\stackrel{\rm cp}{\to}&  
\frac{1}{(2\,\pi)^2}\,\lcbr {{e^{-\Delta_{\rm IR}}}\over {m_im_j}}
\rcbr^{-\alpha A_{ij}}\,\frac{1}{\Gamma^2(\alpha A_{ij})}  \nl
{} &{}& {}  \nl
{} &\times& \int_0^{\infty}\, d\sigma d\sigma'\, 
\lpar \sigma\sigma'\rpar^{\alpha A_{ij} - 1}\, \delta^4\lpar
\sigma p_i + \sigma' p_j - K\rpar.
\eqa
This results explains why we have introduced the term {\em coplanar}.
Note that $\alpha A \sim \beta$ only when the corresponding 
invariant is much larger than mass${}^2$ but the above expression is valid 
for all regimes and it is easily generalized to $n$ emitters
with the result that~\footnote{A.Ballestrero, G.P. work in progress} 
in a process $2 \to n$ any external charged leg $i$ talks to all other 
charged legs, each time with a known scale $s_{ij}$ and with a known total 
weight proportional to
\bqa
x_i^{\alpha\,\lpar A^i_1 + \dots + A^i_I \rpar -1}
\,/\, \Gamma\lpar \alpha\,\lpar A^i_1 + \dots + A^i_I \rpar \rpar, 
\qquad 0 \le x_i \le 1
\eqa
Note that each $A$ has the appropriate sign, in/out, part/antp.
Furthermore, $I(i)$ is the number of pairs $<ij>$ with $i$ fixed.
The IR exponent is given by
\bqa
\alpha A &=& \frac{2\,\alpha}{\pi}\,\lcbr \frac{1+r^2}{1-r^2}\,\ln\frac{1}{r} -
1\rcbr,  \quad
\frac{\mes}{|t|} = \frac{r}{(1-r)^2}  
\eqa
For Bhabha scattering we will have the following combination:
\bqa
-A(s,\me)-A(t,\me)+A(u,\me) &=& 
\frac{2}{\pi}\,\Big[\ln\frac{st}{\mes u} - 1\Big],
\eqa
obtained as an exact result, not a guess.
\section{Conclusions for QED}

The structure-function language is still applicable but initial state 
structure functions evaluated for one scale is, quite obviously, not enough.
In any process each external leg brings one structure function;
since all charged legs talk to each other, each SF is not function of 
one {\em ad hoc} scale but all $<ij>$ scales enter into SF${}_i$.
The exact spectral-function is a convolution of SF
\bqa
E^{{\rm pair}\,<ij>}(K) &=& \int d^4K' \,\Phi(K')
E^{{\rm pair}\,<ij>}_{\rm cp}(K-K'),  \nl
\Phi(K) &=& \frac{1}{(2\,\pi)^4}\,\int d^4x\,\exp\lcbr i\,\spro{K}{x} +
\alpha\,\lpar I - I_{\rm cp}\rpar\rcbr  \nl
{}&=& \delta(K)+\ord{\alpha}
\eqa
Furthermore, IR-finite reminders and virtual parts can be added according to 
the standard approach of reorganizing the perturbative expansion.

\Acknowledgments

I am grateful to A. Ballestrero for the collaboration in this research and to 
H. Haber for the invitation and for the very pleasant stay at Carmel.

\end{document}